\documentclass[
    superscriptaddress,
    longbibliography,
    twocolumn,
    amsmath,
    amssymb
]{revtex4}

\usepackage{newtxtext}
\usepackage[utf8]{inputenc}
\usepackage[english]{babel}
\usepackage{mleftright}
\usepackage{graphicx}
\usepackage{mathtools}
\usepackage{bm}
\usepackage{xcolor}
\usepackage[version=4]{mhchem}
\usepackage{fancyvrb}
\usepackage{listings}
\usepackage{natbib}
\usepackage{bibunits}
\usepackage{algorithm}
\usepackage{algpseudocode}
\usepackage{amsmath}
\usepackage{xspace}
\usepackage{hyperref}
\hypersetup{hidelinks}
\usepackage{acronym}
\usepackage{comment}
\usepackage{units}
\usepackage{url}

\makeatletter
\def\maketitle{
\@author@finish
\title@column\titleblock@produce
\suppressfloats[t]
}
\makeatother

\algdef{SE}[VARIABLES]{Variables}{EndVariables}
   {\algorithmicvariables}
   {\algorithmicend\ \algorithmicvariables}
\algnewcommand{\algorithmicvariables}{\textbf{Global variables}}

\definecolor{backcolour}{rgb}{0.95,0.95,0.92}
\definecolor{commentgrey}{rgb}{0.4,0.4,0.4}
\definecolor{deepblue}{rgb}{0,0.6,1.0}
\definecolor{lightgrey}{rgb}{0.99,0.99,0.99}
\definecolor{deepred}{rgb}{0.0,0.6,1.0}
\definecolor{deepgreen}{rgb}{1.0,0.0,0.6}

\newcommand{\dx}{\mathrm{d}}
\newcommand{\inserteq}[1]{
 \begin{align}
  #1
 \end{align}
}

\newcommand{\chk}[1]{{\color{red}}}

\begin{document}

\title{Meaningful machine learning models and machine-learned \\pharmacophores from fragment screening campaigns}

\author{Carl Poelking}
\email{carl.poelking@astx.com}
\affiliation{Department of Chemistry, University of Cambridge, UK}
\affiliation{Astex Pharmaceuticals, Cambridge, UK}

\author{Gianni Chessari}
\affiliation{Astex Pharmaceuticals, Cambridge, UK}

\author{Christopher W. Murray}
\affiliation{Astex Pharmaceuticals, Cambridge, UK}

\author{Richard J. Hall}
\affiliation{Astex Pharmaceuticals, Cambridge, UK}

\author{Lucy Colwell}
\affiliation{Department of Chemistry, University of Cambridge, UK}

\author{Marcel L. Verdonk}
\affiliation{Astex Pharmaceuticals, Cambridge, UK}

\begin{abstract}
Machine learning (ML) is widely used in drug discovery to train models that predict protein-ligand binding. These models are of great value to medicinal chemists, in particular if they provide case-specific insight into the physical interactions that drive the binding process. In this study we derive ML models from over 50 fragment-screening campaigns to introduce two important elements that we believe are absent in most -- if not all -- ML studies of this type reported to date: First, alongside the observed hits we use to train our models, we incorporate true misses and show that these experimentally validated negative data are of significant importance to the quality of the derived models. Second, we provide a physically interpretable and verifiable representation of what the ML model considers important for successful binding. This representation is derived from a straightforward attribution procedure that explains the prediction in terms of the (inter-)action of chemical environments. Critically, we validate the attribution outcome on a large scale against prior annotations made independently by expert molecular modellers. We find good agreement between the key molecular substructures proposed by the ML model and those assigned manually, even when the model's performance in discriminating hits from misses is far from perfect. By projecting the attribution onto predefined interaction prototypes (pharmacophores), we show that ML allows us to formulate simple rules for what drives fragment binding against a target automatically from screening data.
\end{abstract}

\maketitle

\begin{bibunit}

\begin{figure*}[t]
\centering
\includegraphics[width=0.95\linewidth]{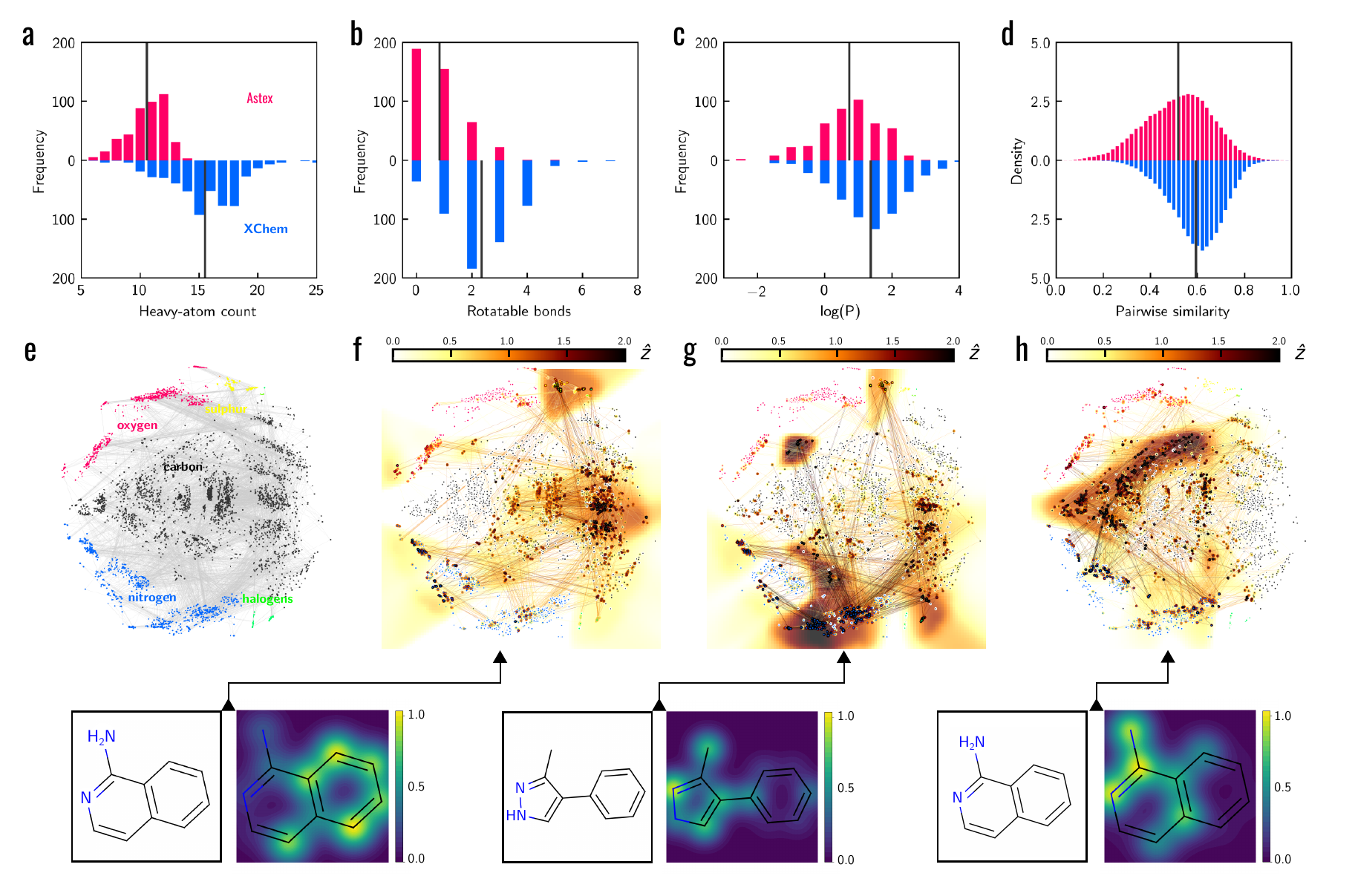}
\caption{
(a-d) Physicochemical profiles of the two fragment libraries associated with the Astex (red) and XChem (blue) datasets: (a) heavy-atom count, (b) number of rotatable bonds, (c) the partition coefficient ClogP, (d) pairwise similarity of chemical environments measured by the $G_{nl}Y_{lm}$ descriptor (see Methods). (e) Chemical map of the Astex fragment library indicating elemental composition and chemical clusters. The colour code is black: carbon, red: oxygen, blue: nitrogen, green: halogens, yellow: sulphur. (f-h) Projections of predicted activity scores onto the chemical map for the (f) global model, (g) kinase model, (h) uPA model (the latter being just one example of a site-specific model). The colour coding ranges from white (neutral) to black (high activity). The bottom row shows the filtered attribution weights $F_\pm(z)$ projected onto fragment hits for the, left, global hit-rate model; centre, kinase hit-rate model; right, active-site model of the serine protease uPA. The colour scale ranges from $F_\pm = 0$ (blue, low significance) to $F_\pm = 1$ (yellow, high significance). }
\label{fig:intro}
\end{figure*}

Fragment-based drug discovery (FBDD) is now a well-established method for obtaining weak, low-molecular weight hits that provide efficient starting points for hit-to-lead optimisation against a protein target of interest~\cite{rees_fragment-based_2004,thomas_structure-guided_2019}. The output from fragment screening campaigns provides a rich source of data for understanding the types of interactions that drive protein-ligand binding for a given protein. Firstly, because of their small size and low complexity, fragment screens tend to provide significantly higher hit rates than screening campaigns involving larger molecules~\cite{jhoti_rule_2013}. Additionally, particularly for well-designed fragment libraries, differences between hits (actives) and misses (inactives) in terms of functional groups or physicochemical profile should be easier to identify than for libraries containing larger, more decorated compounds. However, manually analysing and interpreting the output from a fragment screening campaign of ~1000 fragments can be challenging. The current study shows how the careful use of machine-learning (ML) methods can automate this process.

ML approaches have routinely and successfully assisted the drug discovery workflow for many decades now~\cite{vamathevan_applications_2019}. Over recent years, the rapid evolution and adoption of ML has contributed a large array of novel techniques and ideas~\cite{bajorath_artificial_2020,coley_autonomous_2020}: These include new strategies for modelling of chemically sparse data~\cite{poelking_noisy_2019}, deep representations of molecular structures~\cite{gomez-bombarelli_automatic_2018}, neural-network based docking and scoring functions~\cite{wallach_atomnet_2015} coupled with data augmentation~\cite{scantlebury_data_2020}, data-driven predictions of force-field parameters~\cite{li_machine_2017}, generative modelling~\cite{mendez-lucio_novo_2020}, imputation of assay data~\cite{whitehead_imputation_2019}, etc. Some of these strategies are, however, not without problems~\cite{wallach_most_2018,mccloskey_using_2019}. Indeed, there are perhaps three reasons why the adoption of these exciting technologies may lag behind the astonishing pace at which new techniques are being developed: First, a chronic lack of reliable training data. Second, a chronic lack of reliable models. Third, a chronic lack of trust in the reliability of the (reliable) models. This lack of trust is partly blamed on the black-box character of any sufficiently sophisticated machine-learning framework -- which, as a clich\'e, is partly true, and partly outdated: Tremendous effort has been made over recent years to improve the interpretability of machine-learned models through improved attribution techniques and visualization~\cite{sundararajan_axiomatic_2017}. 

In a drug discovery context, attribution involves mapping the predicted output (such as a binding affinity) back onto features of the input (the molecular structure). The complexity of the attribution procedure grows with the complexity of the underlying machine-learning model. Still, even for simple molecular-fingerprint-based methods, deciding on the quality of this attribution can be a contentious issue, as disagreement between models is commonplace, and the ``true'' answer to the attribution problem is typically unknown~\cite{sheridan_interpretation_2019,mccloskey_using_2019}. In particular, Sheridan observed that standard ML architectures (in his case, different combinations of fingerprints and classifiers) routinely disagree in their attributions, despite agreeing with respect to the predicted activities~\cite{sheridan_interpretation_2019}. Using synthetic data, McCloskey {\it et al$.$} furthermore noted that ML classifiers may exist in a state of deceptive bliss, where the predictions as measured by standard metrics appear close to perfect, but the inferred binding logic has severe deficiencies~\cite{mccloskey_using_2019}. Finally, Sundar {\it et al$.$} proposed decoys as a means to avoid attribution false negatives, and ascribed attribution false positives to background correlations caused by finite sampling. Still, in the development and validation of new attribution procedures, these studies share a common hurdle -- namely, the lack of a ground truth derived from realistic experimental rather than synthetic data. Being able to assess and compare models based on this more fine-grained ground truth rather than on binary binding labels alone would be of great benefit by enforcing chemical realism and thus reducing false positives in virtual screening campaigns~\cite{adeshina_machine_2020}.

For FBDD in particular, interpretation through attribution is desirable because it enables medicinal chemists to make sense of the output of a fragment-screening campaign; pointing, for example, to a specific set of interactions that need to be preserved for binding to occur. In this paper we show that identifying the "hot" regions within fragment hits can be achieved remarkably well with ML methods if sensible attribution strategies are paired with carefully balanced datasets incorporating ``true'' hits and ``true'' misses. We use data from over 50 fragment screening campaigns, both from internal Astex projects as well as from fragment screens performed at the Diamond Light Source to train ML models designed to distinguish hits from misses. For the Astex internal datasets the misses are experimentally validated inactives, which as we illustrate enhances the ML method's ability to provide accurate interpretable models. The attribution procedure becomes robust and virtually unambiguous by resorting to an atomic-environment-based description of the molecular structure, kernel learning and simple filtering. We thus identify which regions in the fragments drive binding against each target. Crucially, we verify that the attribution of the ML binding models is not only consistent internally, but also with expert judgement. The validation is thus approached from two angles using, first, a metric for the spatial {\it co-localization} of attribution weights in the context of the binding site; second, a large number of manual annotations that preceded the modelling. The analysis shows that the ML models are able to successfully identify the "hot" regions within fragments across a wide range of targets. 

\section*{Results \& Discussion}
\label{sec:rd}
\subsection*{Datasets}
\label{subsec:ds}

\begin{figure*}[t]
\centering
\includegraphics[width=0.9\linewidth]{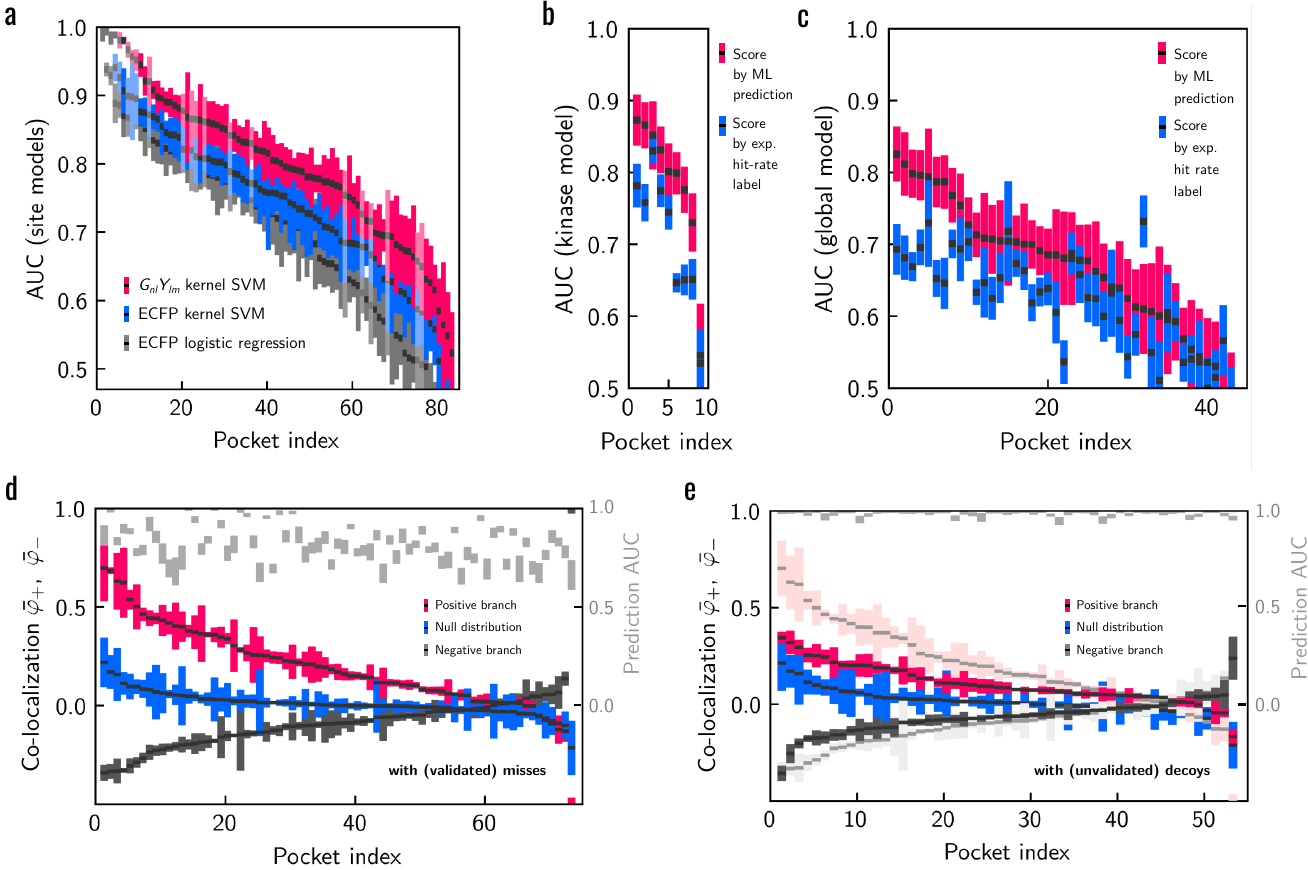}
\caption{
    AUC statistics aggregating the results from the Astex and XChem data for the (a) site-specific models, (b) kinase model, (c) global model. The red bars indicate the standard error of the AUC for each test case. The blue bars in (b) and (c) correspond to AUCs measured for the input labels $y_{hr}^A$ rather than predicted scores $z_{hr}^A$ (see main text for discussion). Lighter shading indicates results obtained for the XChem set. The pocket index enumerates the different pockets across all targets for which independent models were trained, sorted here by the test prediction AUC. For the kinase model, only the pockets corresponding to the active site are considered. (d) Autocorrelation metrics $\bar{\varphi}_\pm$ capturing non-random spatial correlations of the attribution weights across the superimposed fragment hit densities. Red, grey and blue bars correspond to: $\bar{\varphi}_+$ in decreasing order, $\bar{\varphi}_-$ in increasing order, and the null metric $\bar{\varphi}_{0+}$, respectively. (e) Autocorrelation metrics for models derived from validated hits augmented with synthetic decoys (saturated colours) compared to models derived from validated hits {\it and} misses (pale curves, as reproduced from panel (d) for systems where validated misses were available). Note the grey bars in panels (d) and (e): These indicate the AUCs of the pockets when sorted in accordance with the co-localization metric of the positive (red) branch, and illustrating the vanishing correlation between the prediction and attribution performance. }
\label{fig:auc}
\end{figure*}

The fragment binding data we use here stems from two sources: The {\it Astex dataset}, encompassing \chk{} 52 different protein targets from Astex projects; and the {\it XChem dataset}, consisting of 26 targets~\cite{web:xchem, web:fragalysis}. Both datasets are derived from the direct use of X-ray crystallography in a screening mode where a library of diverse fragments were either soaked into protein crystals or co-crystallised with the protein. Within the Astex data, a fragment is designated a ``validated hit'' only if clear and unambigous electron density could be assigned to it. For the analysis we define as the {\it Astex fragment library} the union of all the fragments that at some stage formed part of the (evolving) Astex {\it X-ray} fragment library, whereas the {\it XChem fragment library} was approximated via the union of all reported fragment hits across the different screening campaigns (see Methods section). After postprocessing we obtain 1660 validated hits (26900 misses) for the Astex, and 690 validated hits (18200 misses) for the XChem dataset. Next to site-specific models that consider individual binding pockets or allosteric sites on a protein target, we also construct global hit-rate and kinase hit-rate models that are trained on data aggregated across a larger set of proteins (see Methods section for details).

Fragments are typically selected to be small yet diverse, contain functional groups representative of drug-like compounds, have high aqueous solubility and desirable physical properties such as low lipophilicity. To build intuition for the library composition and the relative frequency of different chemical environments, we include in Fig.~\ref{fig:intro} visualizations of the physicochemical and structural space covered by the fragment libraries. The distributions of heavy-atom counts, rotatable bonds and ClogP (calculated logP, a measure of lipophilicity) in Fig.~\ref{fig:intro}a-c indicate that the XChem fragment library (blue histograms) involves somewhat larger and more complex molecules than the Astex fragment library (red histograms). The distribution of pairwise similarities $k_{ab} = \bm{x}_a \cdot \bm{x}_b$ in Fig.~\ref{fig:intro}d ($\bm{x}_a$, $\bm{x}_b$ are descriptor vectors of atomic environments $a,b$, see Methods) highlights the large diversity of both and, in particular, the Astex set on the sub-fragment level: Fig.~\ref{fig:intro}e elaborates on this further using a low-dimensional projection of the atomic environments that are part of the Astex fragment library, obtained by approximately reproducing the distance $d_{ab} = \sqrt{1 - k_{ab}}$ in the 2D plane using a harmonic-network optimization. Each point of the projection represents an atom-centred chemical environment and is coloured according to the atomic element of that atom, with the lines in this plot representing covalent bonds between atoms. Some major clusters can be assigned to: aliphatic and aromatic carbon environments (centre left and centre right), carbon in aromatic heterocycles (centre), amines, amidines and conjugated nitrogen (bottom left), and acids (top left). We will return to these maps to elucidate the activity of chemical environments on a coarse-grained level.\\

\subsection*{Fragment binding models}
\label{subsec:fbm}

We briefly summarize the key outputs of our ML approach (see the Methods section for details). Once trained on fragment-binding data of validated hits and misses, our ML model equips us with: a local environment-based and global molecular similarity measure (kernel); an overall binding score $Z_A = \sum_{a \in A} z_a$ for a molecule $A$ with atomic-environment-based attributions $z_a$; and a predicted overall binding label $y_A = \mathrm{sign}(Z_A)$ (to be compared against the true experimental label $y_A^\mathrm{true}$).

One of the aims of this paper is to rigorously assess the physical realism of the atom-centred attribution weights $z_a$. First, however, we verify that the individual binding models are able to adequately classify test compounds into hits and misses. We consider three different model types: Site models, which are derived for a particular pocket on a particular target; kinase (class) models, trained by aggregating the sets of hits and misses across the active sites of multiple kinase targets; and global models, trained by aggregating hits and misses across multiple targets and pockets, whatever their type or identity.

The classification results from the cross-validation experiments are summarized in Fig.~\ref{fig:auc}. The testing protocol for the site-specific models is based on random cross-validation (CV) with a training fraction of 0.7: This validation mode can be problematic if the set of hits includes many close analogues of the same warhead, hence exposing the models to structural bias. Fragment libraries, however, are designed around low-MW compounds sampled from a diverse, unbiased set. Even simple CV thus gives an adequate estimate of a model's predictive power. We use the area under the receiver operating characteristic (AUC) as performance metric. The results for the individual pockets from the Astex set (see Fig.~\ref{fig:auc}a, dark colours) range between AUCs of $0.92$ down to $0.55$. We note that some of the models trained on the XChem data (light colours) achieve higher AUCs, due to the less challenging composition of those datasets (which can include analogues) and lack of validated misses. The average AUC of around $0.75$ is expectedly low given the challenging composition (and, interestingly, increases considerably as true negatives are replaced by decoys, see below).

\begin{figure*}[t]
\centering
\includegraphics[width=0.9\linewidth]{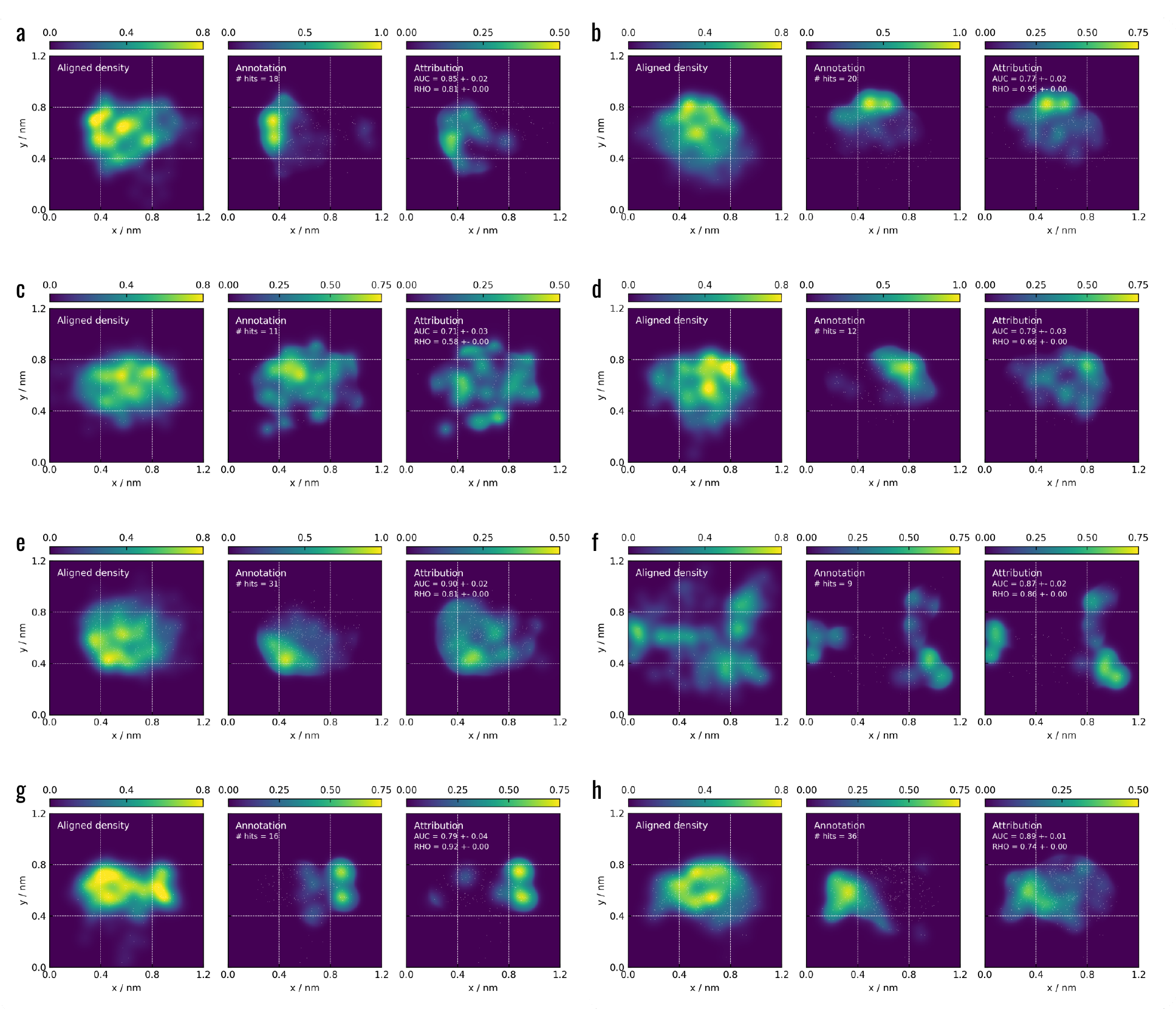}
\caption{ Comparison between manual and machine-learned annotations of superimposed fragment hits for eight systems (a-h). Each triptych consists of: left, the superimposed ligand density; centre, the manually assigned pharmacophore weights mapped onto this density; right, the attribution field derived from the filtered (ranked) weights. The superposition was obtained by aligning the binding sites of the X-ray structures of the protein-ligand complexes. All densities are visualised on the plane of best fit corresponding to the first two principal components of the nuclear coordinates. }
\label{fig:juxta}
\end{figure*}

For the kinase class model (Fig.~\ref{fig:auc}b), we adapt the CV procedure to mimic more closely the situation of a prospective screen against a new kinase target: We derive the classification of fragments into low- vs high-hit-rate compounds from all but one of the kinase datasets and then train a model on these data. Finally we use the {\it hit-rate} labels $\{ y^{hr}_A \}$ (separating high $=+1$ from low $=-1$ hit rates, see Methods) and hit-rate scores $\{ Z^{hr}_A \}$ predicted for the kinase left out in the training as a proxy for the {\it binding} labels $\{ y_A \}$ and scores $\{ Z_A \}$, respectively. The resulting AUCs, with each kinase target left out in turn, are shown in Fig.~\ref{fig:auc}b (red bars): They are surprisingly high, with all but two systems achieving a value larger than $0.75$. Intriguingly, the AUCs obtained by ranking the compounds by the predicted scores $\{ Z^{hr}_A \}$ are better than if we rank them by the true hit-rate labels $\{ y^{hr}_A \}$ on which the models were trained (blue bars). This is a first indicator that the machine-learning model is able to identify relevant subpatterns within the molecular structures to a degree that goes beyond global molecular recognition or scaffold recall~\cite{wallach_most_2018}. Similar conclusions hold for the global model (Fig.~\ref{fig:auc}c), which is evaluated analogously.

Returning to the site-specific models, we note that we also trained baseline models derived from topological molecular fingerprints used in either a kernel setting or logistic regression (see Methods for details). Whereas the topological kernel has a slight edge over the logistic regression, the measured AUCs are nevertheless on average by $0.05$ (or approximately $8\%$) lower than those achieved by the convolutional best-match $G_n Y_{lm}$ kernel described in the Methods section. The latter is therefore attractive both regarding the prediction performance as well as the uniqueness and ease of the attribution procedure. This procedure is somewhat more involved and requires, e.g., iterative dropout applied to molecular substructures and/or input features when global topological fingerprints are paired with nonlinear classifiers such as neural networks or logistic layers~\cite{sundar_attribution_2020,sheridan_interpretation_2019}.

\subsection*{Attribution validation} 
\label{subsec:attr}

The aim of this section is to achieve a statistical validation of the attribution procedure that goes beyond an anecdotal case-by-case confirmation of individual attribution outcomes.

By projecting the whitened (i.e., centered and scaled) atomic attribution weights $\hat{z}_a$ (see Methods, Eq.~\ref{eq:attr}) onto the chemical maps in Fig.~\ref{fig:intro}, we gain a first coarse-grained understanding of which regions of chemical space a model looks at most in arriving at its predictions. For the global and kinase models (Fig.~\ref{fig:intro}f and g) a fairly clear picture results. Globally (and expectedly) we observe a strong lipophilic hotspot (in black) corresponding to aromatic carbon environments (centre right) and sulphurs (top right). Aliphatic carbons on the other hand remain almost surprisingly quiet: Notice the absence of the attribution signal on the left-hand side of the carbon domain, which also reflects the higher hit-rate achieved by flat structures. For kinases, nitrogen groups (in particular, aromatic nitrogen, center bottom) dominate. Finally, for the site-specific model (Fig.~\ref{fig:intro}h, corresponding to the active site of the serine protease uPA in this case), the activity cannot be reduced to a single chemical cluster. The attribution, however, still points to a simple underlying pharmacophore, as exemplified by the highlighted amino-isoquinoline.

On a case-by-case basis such a more detailed picture is arrived at by projecting the attribution weights $z_a$ onto individual molecular structures. Prior to this projection, an additional filtering step maps $z_a$ onto an attribution signal $F(z_a) \in [0,1]$, with larger values of $F(z)$ indicating higher confidence that the assigned weight is statistically significant. In Fig.~\ref{fig:intro}h, we thus see for a particular fragment hit against uPA how the model highlights a feature on the 2-amino-pyridine substructure, which, indeed, forms a salt bridge with the protein. For the kinase model (panel g), the pyrazole of a phenylpyrazole derivative is highlighted, again perfectly in line with chemical intuition. For the global model (panel f) we consider the same isoquinoline derivative as tested against uPA; this time, however, the lipophilic region complementary to the salt-bridge-forming amidine lights up, reproducing the fact that in general lipophiles tend to achieve higher hit rates.\\

\textbf{Co-localization metric.} For a first more comprehensive validation of our attribution procedure, we use the relative positioning of the ligands within their binding site as derived from spatially aligned X-ray structures: The underlying assumption is that hotspots on the ligands correspond to hotspots on the protein. Environments with large attribution weights, when mapped onto the superimposed X-ray binding modes of the fragments, should thus cluster (i.e., ``co-localize'') in space. This information -- which is {\it not} made available to the ML models during training -- leads to a superposition density onto which we can subsequently project the attribution weights. The validation step then consists of evaluating a spatial autocorrelation statistic $|\bar{\varphi}_\pm| \leq 1$ of the attribution weights relative to their local attribution ``field'' (see Methods for the definition of $\bar{\varphi}_\pm$). An autocorrelation with a magnitude larger than a random baseline indicates that the attribution procedure was successful in the sense that chemical environments deemed significant to the binding tend to localize in the same region of the binding site.

Fig.~\ref{fig:auc}d summarizes the co-localization metrics $\bar{\varphi}_\pm$ calculated for a range of binding sites. To distinguish between the prediction of ``hot'' and ``cold'' spots, $\bar{\varphi}_+$, indicated by the red bars, measures the co-localization of chemical environments with {\it positive} (higher than average) weights; $\bar{\varphi}_-$, indicated by the grey bars, measures co-localization of environments with {\it negative} (lower than average) weights. The blue bars indicate the null background $\bar{\varphi}_{0}$ derived from scrambled (randomly permuted) weights. The comparison with this null background highlights that the attribution weights exhibit significant non-random spatial clustering. The autocorrelation of the weights is more pronounced for the positive branch $\bar{\varphi}_+$ than the negative branch $\bar{\varphi}_-$, indicating that the hotspots do indeed cluster significantly better in space than the predicted cold spots. This is not surprising given that the autocorrelations are computed only across fragment hits, which as such exclude clashes that could result in large negative attribution weights.

To understand better the role played by the {\it inactive} data (i.e., validated misses) in informing the attribution model, we repeat this analysis for synthetic datasets that combine validated hits from the Astex sets (for which we have true validated misses) with unvalidated {\it decoys} sampled from the XChem fragment library. As shown in Fig.~\ref{fig:auc}e, this replacement of misses with decoys results in a noticeable decay of the autocorrelation signal, whereas the prediction AUCs increase drastically to almost unity (see the pale grey bars at the top of this plot compared to those in panel d): This shows that the models get overconfident in forming their predictions, while the attribution becomes less specific and smeared out -- a finding that agrees well with previous studies that observed that large AUCs are not necessarily a good indicator of chemical soundness~\cite{mccloskey_using_2019,wallach_most_2018}. Even for the original datasets that incorporate both validated hits and misses, the correlation between the autocorrelation metric and AUCs is marginal at best (see the red bars and pale grey bars at the top of Fig.~\ref{fig:auc}d, respectively). This disconnect between classification and attribution performance reminds us that we are still in a sampling regime where the inferred fragment logic does not have to be complete to arrive at (coincidentally) correct predictions. True negative data then acts as a natural filter that forces the models to assign atomically ``sparse'' and meaningful attributions.\\

\begin{figure*}[t]
\centering
\includegraphics[width=0.9\linewidth]{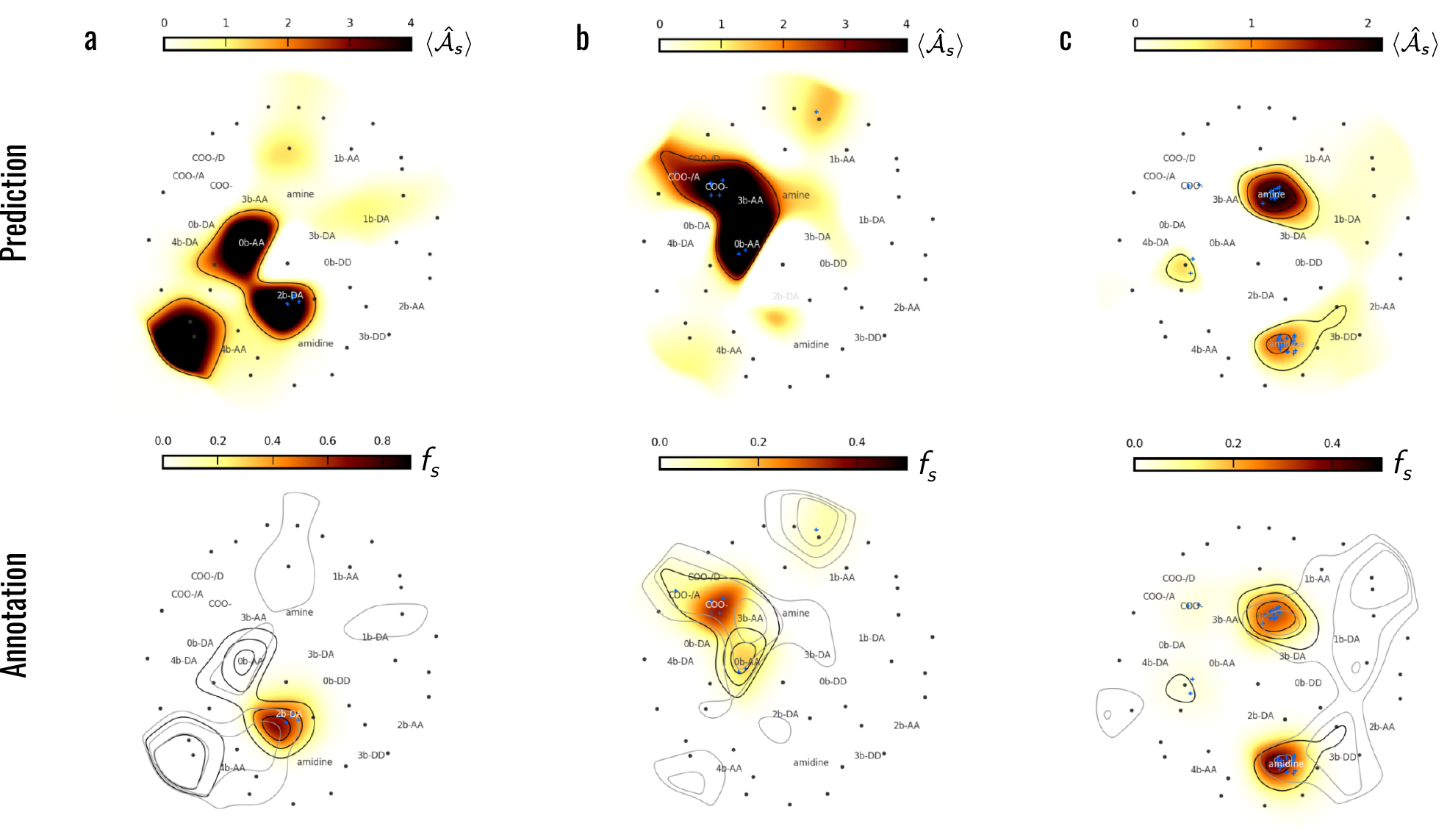}
\caption{Projections of machine-learned (top row) and manually assigned (bottom row) pharmacophore amplitudes onto two-dimensional pharmacophore maps derived from co-occurrence statistics. The projections are obtained from the site-specific models for (a) the ATP site of the kinase CDK2 (b) the protein-protein interaction target KEAP1, (c) the trypsin-like serine protease uPA. The colour maps for the top row indicate the locally averaged pharmacophore amplitude $\hat{\mathcal{A}}$ (Eq.~\ref{eq:phcore_amp}). For the bottom row, the colour reflects the relative frequency of a pharmacophore across the set of annotated fragment hits.}
\label{fig:smarts}
\end{figure*}

\textbf{Expert annotations.} The co-localization analysis indicates that the attribution procedure is internally consistent, with hotspots inferred by the machine-learning protocol clustering together within the binding site. We now complement this validation route with a yet more stringent test, comparing the derived hotspots with those annotated manually by human experts. These annotations, performed prior to any machine learning, involved assigning a {\it minimal pharmacophore} to each fragment hit by visually inspecting the fragment-protein complex. The minimal pharmacophore of a fragment hit reflects what interaction motif drives the binding of the fragment to the protein and which atoms in the fragment are involved in these interactions. Examples of minimal pharmacophores include a donor-acceptor pattern in the fragment forming a pair of hydrogen bonds with the hinge region of a kinase target, or an amidine group forming an ion pair with the aspartic acid in the S1 pocket of a serine protease. The definitions of these minimal pharmacophores are stored as SMARTS patterns that are mapped onto the chemical structures. We denote this mapping as a spin vector for each molecule with components $m_a = \pm 1$ that indicate whether an atom $a \in A$ forms part of the pharmacophore ($m_a = +1$) or not ($m_a = -1$). 

To assess the degree to which the annotations $\{ m_a \}$ agree with our attribution weights, we map both onto the superimposed point densities of the fragments and project the resulting fields onto the plane of best fit of the ligand density. Fig.~\ref{fig:juxta} shows a visual comparison of heat map representations of these fields for a subset of eight binding sites, juxtaposed with the superimposed densities. For a quantitative comparison, we evaluate the Pearson correlation $\rho$ between the annotation and attribution fields. This correlation is surprisingly large, ranging between $\rho = 0.58$ to $\rho = 0.95$, and, again, only weakly correlated with the classification metrics achieved by the machine learning: E.g., the site model with the largest observed $\rho = 0.95$ has an AUC of $0.77$, whereas the uPA model achieved an AUC of 0.89, but a field correlation $\rho$ of only $0.74$. This can be rationalized in that simple, but potentially non-specific pharmacophores (such as an acid group) tend to lead to good agreement between the manual and automatic annotation, but poor classification. More precisely, due to the low specificity of the pharmacophore, but also the difficulty of predicting steric clashes (among others), the model has trouble resolving why some fragments featuring the same pharmacophore were reported as a validated miss. The human experts on the other hand may not face the same conundrum, as their annotations are not always (and not easily) checked manually for consistency with the set of misses. This in turn offers another perspective how the attribution output can be combined with manual annotations to check for weakly specific or incomplete pharmacophore assignments. Nevertheless, already in the present form, both the quantitative and visual comparison indicate widespread agreement between the machine-learned and manual attributions both for strongly localized, simple pharmacophores (such as the acid group in Fig.~\ref{fig:juxta}g) as well as more diffuse pharmacophores (such as those of Fig.~\ref{fig:juxta}f).

A potentially problematic issue concerns the degree to which the machine-learned attribution localizes the inferred binding pattern within a given fragment hit. This issue is partly connected to how we originally defined the chemical environments: They are atom-centred, and have a size that is determined by a radial cutoff of, here, \unit[5.5]{\AA}. This implies that, given a large attribution weight, in principle any structural feature contained within the cutoff sphere of an atom could be deemed significant. Fortunately, in practice the attribution tends to localize important groups with a higher resolution than this cutoff radius. We attribute this to the concerted action of the smooth atomic kernel, the competitive matching procedure performed on the molecular level, and post-processing of the weights. Amplified further by effective decoys (ideally, validated misses), the localization can be remarkably strong -- see, for example, the large weight attributed to the covalently bound acetyl group in the case of the SARS-CoV-2 main protease (see Fig.~S2 of the SI appendix)~\cite{douangamath_crystallographic_2020}.

\subsection*{Pharmacophore projections} The analysis of the machine-learned attribution fields illustrates that the attribution procedure often works remarkably well. We now address, as a perspective, whether we can derive explicit (human-readable) pharmacophore trends that go beyond case-by-case visualizations such as those shown in Fig.~\ref{fig:intro}f-h. The idea is to map the attribution weights onto predefined SMARTS patterns representing a variety of potential minimal pharmacophores such as two-bond donor-acceptors, acids, lipophiles, etc. 

The protocol is as follows: First, for a fragment candidate $A$ predicted to bind to the active site of a protein target, we derive the attribution weight vector $\bm{z}$ with components $z_{a}$ (see Methods). Second, we evaluate for each pharmacophore candidate $s$, a vector $\bm{m}_s$ with components $m_{sa} = \pm 1$ that indicate whether or not an atom $a \in A$ participates in pattern $s$. Third, we rank the different patterns $s$ using our weights $\bm{z}_A$ according to an AUC $\mathcal{A}_{s} = \mathcal{A}(\bm{z}, \bm{m}_s)$. In other words, we interpret the attribution weights as scores for classifying individual atomic neighbourhoods into {\it participating} ($m_{sa} = +1$) or {\it non-participating} ($m_{sa} = -1$) environments. See the Methods section for details.

To visualize the outcome, we project the whitened amplitudes $\hat{\mathcal{A}}_s$ onto a two-dimensional map of the minimal pharmacophores. This map is derived from the co-occurrence statistics of these pharmacophores across the fragment library. The ``distance'' between pharmacophores $s$ and $s'$ is evaluated as $D_{ss'} = 1 - C_{ss'}$, with a correlation $C_{ss'} = \sum_{A} I_{As} I_{As'} / \sqrt{N_s N_{s'}}$: Here $I_{As} \in \{0, 1\}$ is an indicator function for the presence of pharmacophore $s$ in fragment $A$; $N_s = \sum_{A} I_{As}$ are total pharmacophore counts.

Fig.~\ref{fig:smarts} exemplifies the resulting pharmacophore heat maps for three systems: (a) the ATP site of cyclin-dependent kinase 2 (CDK2)~\cite{wyatt_identification_2008}, (b) the protein-protein interaction target KEAP1~\cite{heightman_structureactivity_2019}, (c) the trypsin-like serine protease, urokinase plasminogen activator uPA~\cite{frederickson_fragment-based_2008}. As can be seen from the heat maps in the top row, the pharmacophore map displays strong, distinctive hotspots for each system, corresponding to (a) donor-acceptor, (b) acid and acceptor-acceptor, (c) amine and amidine patterns. For comparison, the bottom row of Fig.~\ref{fig:smarts} shows the heat maps derived from the manual pharmacophore annotations: These are clearly more localized and specific than the machine-learned projections. The agreement is adequate, taking into account that some cross-talk between different pharmacophores is to be expected, as the SMARTS definitions are not mutually exclusive: This confusion is particularly visible for the CDK2 kinase in Fig.~\ref{fig:smarts}a. Nevertheless, pharmacophore projections of this type may prove useful for automatic post-processing and analysis of rapid as well as X-ray screening campaigns.\\

\section*{Conclusions}
\label{sec:co}

Even though various studies have investigated the potential of machine-learned models of ligand-protein interactions for drug discovery, attempts to query, assess and visualize the physical interactions learnt and predicted by those models have remained relatively scarce. Here we have validated rigorously and on a large scale the physical binding patterns inferred by state-of-the-art machine learning. Measuring and controlling for both internal consistency and agreement with human experts, our analysis of the attribution outcomes indicates that, overall, the attribution is surprisingly successful in identifying meaningful binding patterns from fragment binding data, even in cases where the raw classification performance is relatively low (AUC < $0.75$). 

The fact that these models are derived from X-ray hits and misses rather than from binding data on larger more complex molecules improves the machine-learning outcome: The quality of the models benefits greatly from a balanced, diverse fragment library with minimal structural redundancy -- design criteria which, if violated, result in models displaying inflated performance measures and an impaired physical ``understanding''. 

The models may in turn help us improve the composition of fragment libraries through global and class-specific hit-rate modelling. Furthermore, and perhaps most importantly, they provide medicinal chemists and modellers with an unbiased and complete view of the output of a fragment screening campaign, e.g., in the form of hot spots within fragment hits, which can be used to guide subsequent fragment-to-lead optimisation.

{\footnotesize

\section*{Methods}

\textbf*{Data preparation.} We consider all crystallographic experiments performed at Astex that involved the soaking or co-crystallisation of a fragment of the Astex X-ray screening fragment library (here referred to as the {\it Astex fragment library}) that forms a subset of Astex's larger biophysical screening fragment libraries. The vast majority of these crystallographic experiments would have formed part of a primary X-ray fragment screen. This {\it Astex fragment library} comprises 1660 validated X-ray hits and \chk{} 26900 validated X-ray misses against primary and allosteric sites, identified by clustering the hits based on distance- and density-based criteria. This results in an average of \chk{} 2-3, and a maximum of up to \chk{} twelve sites per target. For the {\it XChem dataset}, only the hits have been reported. Therefore, we approximated the {\it XChem fragment library} as the union of all hits against all targets in the {\it XChem dataset}. To define the misses for each target, we formed the difference between this {\it XChem fragment library} and the observed hits for that target. Hence, for the {\it XChem dataset} we cannot be certain that all the misses are true misses. \\

\textbf{Hit-rate models.}
Global and target-class datasets are constructed from the site-specific data. For the global model, we divide the compounds of the fragment library into high vs low-hit-rate compounds. As dividing criterion we require high-hit-rate fragments to have resulted in a hit against at least two targets of a {\it different class}, as judged by their EC (enzyme commission) number. This criterion translates into a hit rate of $3\%$ or higher. For the class-specific models, only kinases (EC2.7) were represented among the set of targets with a critical number large enough to warrant modelling.\\

\textbf*{Machine learning.}
The machine-learning framework is based on a local description of atom-centered chemical environments used as input for kernel-based learning and closely related to existing approaches~\cite{bartok_machine_2017,de_comparing_2016}. The model implicitly learns the correlations and co-occurrences of a particular set of atomic environments that set active compounds (hits) apart from inactive compounds (misses). The role of the attribution is then to identify and visualize which correlations and types of atomic environments are key to successful binding. Formulating the predictions in terms of atomic environments greatly simplifies the attribution procedure as is shown further below.

Clearly there is significant freedom in how to choose the atomic description~\cite{parsaeifard_assessment_2020}, ranging from tensor-based methods~\cite{shapeev_moment_2016}, graph-convolutional neural networks~\cite{kearnes_molecular_2016}, to hard-coded convolutional descriptors such as atom-centred symmetry functions and SOAP (Smooth Overlap of Atomic Positions~\cite{behler_constructing_2015,bartok_representing_2013}). In this work, all we require is that we can construct a smooth but nevertheless sensitive similarity measure from the atomic description. Here we therefore follow a simple approach where the atomic descriptor vector $\bm{x}^a$ is obtained from a basis-set expansion $\bm{c}^a$ of the local neighbourhood of a heavy atom $a$ in molecule $A$,
\inserteq{
 c^a_{t n l m} &= \sum_{a' \in A} w^{a'}_t f(r_{aa'}) G_{nl}(r_{aa'}) Y_{lm}(\bm{r}_{aa'}),
}
where $w^{a'}_t$ are weights indicating the type $t$ of atom $a'$ (carbon, hydrogen, etc.), $f(r)$ is a cutoff function, $G_{nl}(r)$ are radial basis functions of index $n$ and angular momentum $l$, and $Y_{lm}(\bm{r}_{aa'})$ are real spherical harmonics. The radial basis functions are in turn constructed from Gaussian functions $g_n(r)$ modulated by a frequency-damping function $\lambda_l(r)$ that suppresses high-$l$ components at shorter distances, and a radial decay $h(r)$ that approaches $r^{-2}$ for large $r$ and thus discounts distant neighbours:
\inserteq{
 G_{nl}(r_{aa'}) &= g_{n}(r_{aa'}) h(r_{aa'}) \lambda_l(r_{aa'}), \\
 \lambda_l(r_{aa'}) &= \exp\left( -\sqrt{\frac{2 l}{\pi}} \frac{\sigma}{r_{aa'}} \right).
}
Here $\sigma = \unit[0.5]{\AA}$ is an atomic-width parameter that enforces smoothness. Finally, the atomic descriptor is obtained by rotational averaging of the $lm$-components:
\inserteq{
 x^a_{tunkl} = \sum_m c^a_{tnlm} c^a_{uklm}.
}
The above framework, notably the rotational averaging performed at the end, is virtually equivalent to the SOAP formalism~\cite{bartok_representing_2013}, but has the advantage that it is faster to evaluate due to the damping functions $\lambda_l(r)$ and offers better control over longer-range vs short-range contributions due to the distance decay $h(r)$. The convolutions leading to $x^a_{tunkl}$ suggest a ``native'' kernel $k_{ab} = \bm{x}^a \cdot \bm{x}^b / | \bm{x}^a| | \bm{x}^b |$ over atomic environments, which is inherited from SOAP and mimics the rotationally averaged density overlap of two chemical environments.

The molecular representation is constructed implicitly through the choice of kernel function. For intensive properties such as a binding affinity, a best-match procedure~\cite{de_comparing_2016} (as opposed to, e.g., averaging of the atomic descriptor vectors) yields the appropriate molecular kernel:
\inserteq{
 K_{AB} &= \max_p \sum_{a \in A} \sum_{b \in B} k_{ab} p_{ab}, \\
 \mathrm{subject\ to} &\sum_{a} p_{ab} = \frac{1}{N_B} \ \mathrm{and} \ \sum_{b} p_{ab} = \frac{1}{N_A}. \nonumber 
}
The constraints with heavy-atom counts $N_A$ and $N_B$ of molecules $A$ and $B$ enforce competition among the atoms of each fragment for their best-matching partner, and assign to $p_{ab}$ the role of a permutation matrix. This kernel can be easily decomposed onto atomic contributions if used together with support-vector machines or Gaussian processes. The decision function $Z_A$ distinguishing between hits ($z > 0$) and misses ($z < 0$) then reads
\inserteq{
 Z_A = \sum_{B \in \mathcal{T}} K_{AB}^\nu w_B + Z_0,
}
with training set $\mathcal{T}$, kernel power $\nu \geq 1$ and offset $Z_0$. \\

\textbf{Baseline models.} The baseline models use extended-connectivity fingerprints (ECFPs) as implemented by rdkit~\cite{rogers_extended-connectivity_2010,web:rdkit} in combination with kernel support vector machines and logistic regression. As kernel function we choose $k_{AB} = \bm{x}_A \cdot \bm{x}_B / | \bm{x}_A || \bm{x}_B |)^\nu$, with some positive power $\nu \geq 1$. Experience shows that this simple ECFP kernel performs remarkably well in the prediction of molecular properties. The hyperparameters of the baseline models optimized via nested splits are: the bond radius of the ECFP, the kernel exponent $\nu$, and the regularization strength $C$ of the SVM and logistic regression.\\

\textbf{Attribution and filtering.} The attribution of $Z_A$ onto environments $\{ a \}$ is obtained via~\cite{bartok_machine_2017}
\inserteq{
 z_a = \sum_{B \in \mathcal{T}} \left( K_{AB}^{\nu-1} w_B \sum_{b \in B} k_{ab} p_{ab} \right) + \frac{Z_0}{N_A}. \label{eq:attr}
}
One can show (see the SI appendix) that this attribution recipe in fact corresponds to a kernelized version of integrated gradients~\cite{sundararajan_axiomatic_2017}. 

There are several approaches to filtering these weights as to improve their robustness with respect to dataset composition. First, for the comparison with the manual annotations, we use a simple rank filter that, for each molecule, considers only the $n$-largest atomic weights, with $n=3$ in our case. For a case-by-case analysis (see, e.g., Fig.~\ref{fig:intro}, bottom row), it is more useful to filter the atomic weights $z_a$ by magnitude and assign a confidence to each environment $a$ as to whether the sign of $z_a$ is correct: To this end we sample from the training set the distribution $f_-(z)$ for true negatives (aggregating weights from compounds correctly predicted to be a miss) and the distribution $f_+(z)$ for true positives (aggregating weights from compounds correctly predicted to be a hit). Note that these weights are extracted from the training set in a leave-one-out procedure; i.e., the weights contributing to $f_\pm(z)$ are from compounds excluded during the training, under the assumption that the decision function does not change drastically as single hits and misses are left out one by one. Having thus obtained $f_\pm(z)$, we assess negative predicted weights against the cumulative distribution $F_+(z) = \int_{z}^\infty f_+(z') \dx z'$ and positive weights against $F_-(z) = \int_{-\infty}^z f_-(z') \dx z'$. For example, if $F_-(z_a)$ is large (approaching unity), this indicates that the positive weight $z_a$ is abnormally large to have occurred in a true negative compound, and should therefore be considered significant. On the other hand, if $F_-(z_a)$ had a value of close to 0.5, then the magnitude of $z_a$ is in no way extraordinary and could have just as well been observed in a miss.

For the pharmacophore reconstruction, as well as the co-localization metrics, we resort to the z-scored weights as the normalized attributions $\hat{z}_a = (z_a - \mu_z / \sigma_z$: Here $\mu_z$ and $\sigma_z$ are, respectively, the average and standard deviation of the atomic weights predicted for the training set.\\

\textbf{Co-localization metric.}
 We define a attribution correlation field $\varphi(\hat{z})$ as a local distance-weighted average over the attribution weights of the atomic centres of the superposition cloud:
 \inserteq{
  \varphi(\hat{z}) = \left\langle \frac{1}{Q_a} \sum_{B \neq A} \sum_{b \in B} \hat{z}_b \exp(- \alpha r_{ab}^2) \right\rangle_{\hat{z}_a = \hat{z}}.
 }
 Here $Q_a = \sum_{B \neq A} \sum_{b \in B} \exp(- \alpha r_{ab}^2)$ is a normalizing factor; $\hat{z}$ as opposed to $z$ are the whitened (i.e., z-scored) attribution weights. Intuitively, the field $\varphi(\hat{z}_a)$ measures the expected magnitude and sign of the weights in the neighbourhood of an atomic centre $a$ given that its own attribution weight is $\hat{z} = \hat{z}_a$. To obtain the random baseline, we construct a null field $\varphi_0$ by randomly permuting the attribution weights among all atomic centres of the point cloud. Examples for the correlation fields $\varphi(\hat{z})$ and $\varphi_0(\hat{z})$ are included in the SI appendix.
 
 The performance metric for the attribution is subsequently derived as an integral over the positive and negative branches of the correlation field:
 \inserteq{
  \bar{\varphi}_\pm &= \pm \hat{z}_\mathrm{max}^{-1} \int_0^{\pm \hat{z}_\mathrm{max}} \hspace{-0.25cm} \varphi(\hat{z}) \ \dx \hat{z}\ .
 }
 With $z_\mathrm{max}$ chosen as two times the standard deviation of the weights $z$ (i.e., $\hat{z}_\mathrm{max} = 2$), the metrics $\bar{\varphi}_\pm$ assume extremal values of $\pm 1$: These are obtained if the weights of the point cloud are ``perfectly'' correlated in the sense that the background field $\varphi(\hat{z}_a)$ seen by all sites $a$ is identical to their own weight $\hat{z}_a$. In practice, however, even a ``perfect'' model would not be able to achieve $\bar{\varphi}_\pm = \pm 1$, as the boundaries of the hotspots for each fragment hit cannot be expected to align perfectly, and because some hotspots may not be shared by all hits due to different albeit overlapping binding modes.\\

\textbf{Pharmacophore projection.} Due to the small though nevertheless varying size of the fragments and pharmacophores, the SMARTS AUCs are subject to significant statistical fluctuations. We normalize them using the mean $\mathcal{A}_0 = 0.5$ and width $\Delta\mathcal{A}$ of the distribution of AUCs obtained from {\it randomized} instances of the mapping vector $\bm{\tilde{m}}_s$:
\inserteq{
 \hat{\mathcal{A}}_s = \frac{\mathcal{A}(\bm{z}, \bm{m}_s) - \mathcal{A}_0}{\Delta\mathcal{A}(\bm{z}, \tilde{\bm{m}}_s)}.
}
For example, a normalized AUC of $\hat{\mathcal{A}}_s \geq 2$ would indicate strong (``$2\sigma$'') agreement between the attribution weights and the pharmacophore $s$. 

By averaging the whitened AUCs $\hat{\mathcal{A}}_{As}$ over all hits $A \in \mathcal{T}$ of a training set $\mathcal{T}$ of size $N_\mathcal{T}$,
\inserteq{
 \langle \hat{\mathcal{A}}_s \rangle = \frac{1}{N_\mathcal{T}} \sum_A \hat{\mathcal{A}}_{A,s}, \label{eq:phcore_amp}
}
we obtain a measure for the compatibility between pharmacophore $s$ and the fragment hit observations made for a specific binding site.\\

\textbf{Code availability.} The core library implementing the $G_{nl}Y_{lm}$ formalism, kernels and attribution routines is available online at \texttt{github.com/capoe/gylmxx}. Attribution models have been incorporated into the BenchML model library~\cite{poelking_benchml_2021} (see \texttt{github.com/capoe/benchml}). A usage example will be provided in BenchML's {\it examples} folder, located at \texttt{/benchml/examples/fragml}.\\

\textbf{Data availability.} Astex's fragment library is proprietary and can therefore not be made available. The most recent XChem data on the other hand can be obtained at \texttt{fragalysis.diamond.ac.uk}. A curated subset used in this work can be obtained from \texttt{github.com/bingqingcheng/linear-regression-benchmarks} (see the \texttt{binding/xchemfrag} subfolder therein).

}

\end{bibunit}

\clearpage
\setcounter{page}{1}
\setcounter{figure}{0}
\setcounter{section}{0}
\setcounter{equation}{0}
\renewcommand{\thepage}{S\arabic{page}}
\renewcommand{\thesection}{\Alph{section}}
\renewcommand{\thesubsection}{\alph{section}}
\renewcommand{\thetable}{S\arabic{table}}
\renewcommand{\thefigure}{S\arabic{figure}}

\title{SUPPLEMENTARY INFORMATION \\ \vspace{0.5cm} {Meaningful machine learning models and machine-learned \\pharmacophores from fragment screening campaigns}}

\maketitle

\begin{bibunit}

\section*{Connection to integrated gradients}

The attribution weights $z_a$ defined in Eq.~7 of the main text were derived simply and directly from inspecting the decision function $Z_A$. We note here that the same expression can be derived from a kernelized variant of integrated gradients: Sundararajan {\it et al.} proposed the following attribution rule for a function $F$ that maps inputs $x$ onto an output $y$~\cite{sundararajan_axiomatic_2017}:
\begin{align}
 f_i(x) = (x_i - x_i') \int_0^1 \frac{1}{\alpha} \frac{\partial F(x' + \alpha(x-x'))}{\partial x_i} \mathrm{d}\alpha.
\end{align}
Here $f_i(x)$ is the attribution weight assigned to input $x_i$; $x'$ is a baseline input against which the attribution is offset. The factor $1/\alpha$ inside the integral is not part of the original definition, but was added here to ensure the correct attribution of low-order polynomials (noting that for high-dimensional inputs $x_i$ and highly nonlinear functions $F$ the difference is barely noticeable). 

In our kernelized approach, it is natural to consider as inputs the pairwise similarities $k_{ab}$ between the atoms $a$ of a probe structure and the atoms $b$ of fragments $B$ in the training set. We thus associate the attribution weight for atom $a$ with
\begin{align}
 z_a = \sum_{B} \sum_{b \in B} z_{ab}(k) = \sum_B \sum_{b \in B} k_{ab} \int_0^1 \frac{1}{\alpha} \frac{\partial Z_A(\alpha k)}{\partial k_{ab}} \mathrm{d}\alpha,
 \end{align}
where $k$ is the vector of all pairwise kernel values $k_{ab}$. Furthermore accounting for the constant offset $Z_0$ in the decision function (constant terms are ignored by integrated gradients by definition), we thus arrive at the attribution expression of Eq.~7.

\section*{Attribution fields}

 As described in the Methods section, we defined the attribution correlation field $\varphi(\hat{z})$ as a local distance-weighted average over the attribution weights of the atomic centres of the superposition cloud:
 \inserteq{
  \varphi(\hat{z}) = \left\langle \frac{1}{Q_a} \sum_{B \neq A} \sum_{b \in B} \hat{z}_b \exp(- \alpha r_{ab}^2) \right\rangle_{\hat{z}_a = \hat{z}}.
 }
 Here $Q_a = \sum_{B \neq A} \sum_{b \in B} \exp(- \alpha r_{ab}^2)$ is a normalizing factor; $\hat{z}$ as opposed to $z$ are the whitened (i.e., z-scored) attribution weights. Intuitively, the field $\varphi(\hat{z}_a)$ measures the expected magnitude and sign of the weights in the neighbourhood of an atomic centre $a$ given that its own attribution weight is $\hat{z} = \hat{z}_a$. Fig.~S1 exemplifies $\varphi(\hat{z})$ for nine systems, relative to a random baseline, which corresponds to a null field $\varphi_0$ obtained by randomly permuting the attribution weights among all atomic centres of the point cloud.

\section*{Localization of the attribution weights}

The degree to which the machine-learned attribution localizes the inferred binding pattern within a fragment hit can unfortunately not be directly controlled, as the locality is the implicit result of the smooth expansion of the atomic environments and atomic kernel, the competitive matching procedure performed on the molecular level, and post-processing of the weights. 

For an example and visual guide to how the framework localizes the attribution weights, Fig.~S2 shows the superimposed densities and attribution fields for four binding sites taken from the XChem dataset~\cite{web:xchem}. We discuss briefly the SARS-CoV-2 main protease, for which XChem recently reported the results of a fragment-screening campaign. Fig.~S2b shows the superimposed density of the covalently bound ligands on the left together with the machine-learned attribution field on the right: Even though the scaffolds of the fragment hits overlap significantly, the attribution focuses only on the acetyl group that binds covalently to the protein. Naturally this group also turns out to be the one that is spatially best conserved. This information, however, was not made available to the model. The non-covalent hits against the same protease on the other hand present a much more diffuse picture (Fig.~S2c): The fragment hits display significant scattering, which can explain why the machine-learned attributions are unable to pinpoint a single hotspot.\\

\section*{Expert annotations}

Figs.~S3-S5 compare expert annotations, based on minimal-pharmacophore definitions, with machine-learned attributions for a total of eighteen binding sites. Each row in the figures visualizes the outcome for a single site, including, from left to right: the superimposed ligand density, the manually assigned pharmacophore, the attribution weight density, and the standard deviation of the weight density. The visualizations are projected onto the plane of best fit as derived from the principal-component analysis of the nuclear coordinates of the superimposed structures, where the superposition is derived from aligning the binding site of the protein as resolved by X-ray. The standard deviation is calculated by resampling with replacement (i.e., bootstrapping) the points (i.e., atoms) of the superimposed cloud density. For each sample, the field is re-evaluated on a grid. Finally the component-wise standard deviation is calculated over the set of samples.

\end{bibunit}

\begin{figure*}[t]
\centering
\includegraphics[width=0.65\linewidth]{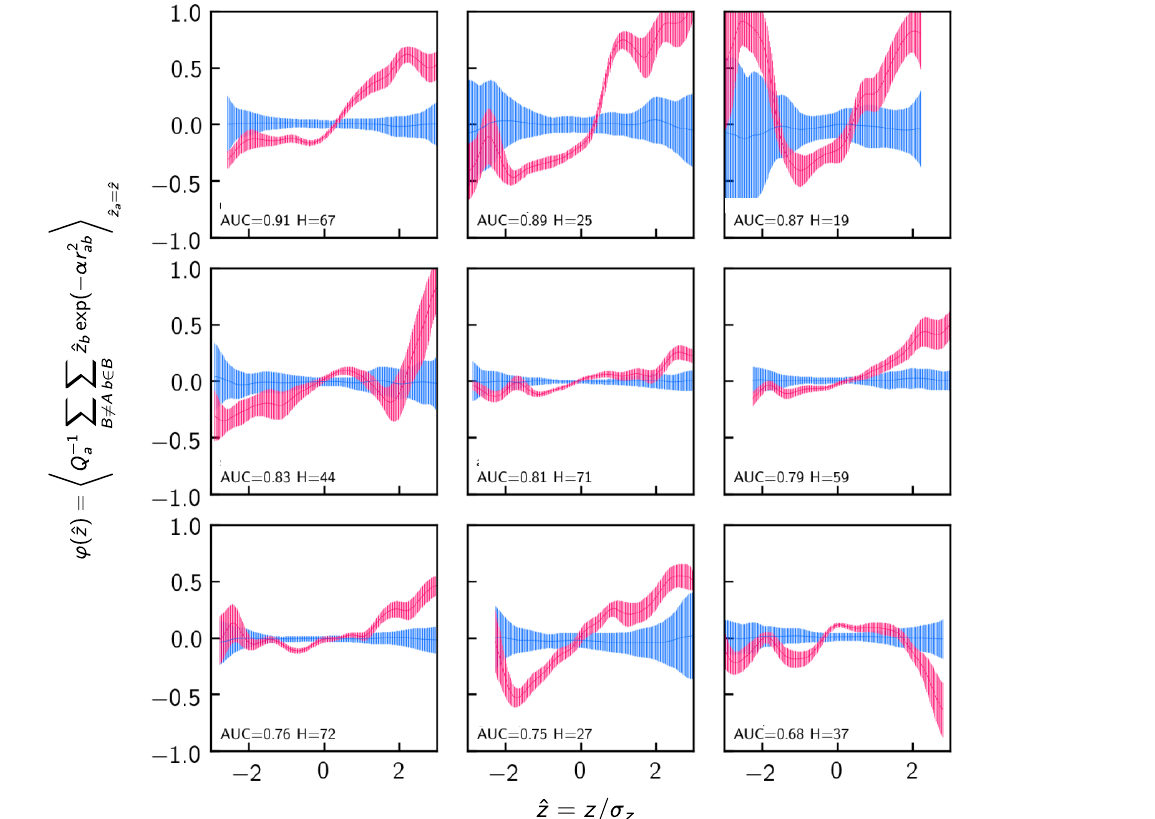}
\caption{ The attribution correlation fields for nine example systems: The red curves show $\varphi(\hat{z})$, with confidence intervals obtained by bootstrapping the set of atoms that form the superimposed ligand density, and re-evaluating $\varphi(\hat{z})$ for every bootstrapped sample. The blue curves represent the null background fields calculated for scrambled data, where the atomic attribution weights were permuted randomly among the atoms of the superimposed density. }
\label{fig:siacorr}
\end{figure*}

\begin{figure*}[t]
\centering
\includegraphics[width=0.65\linewidth]{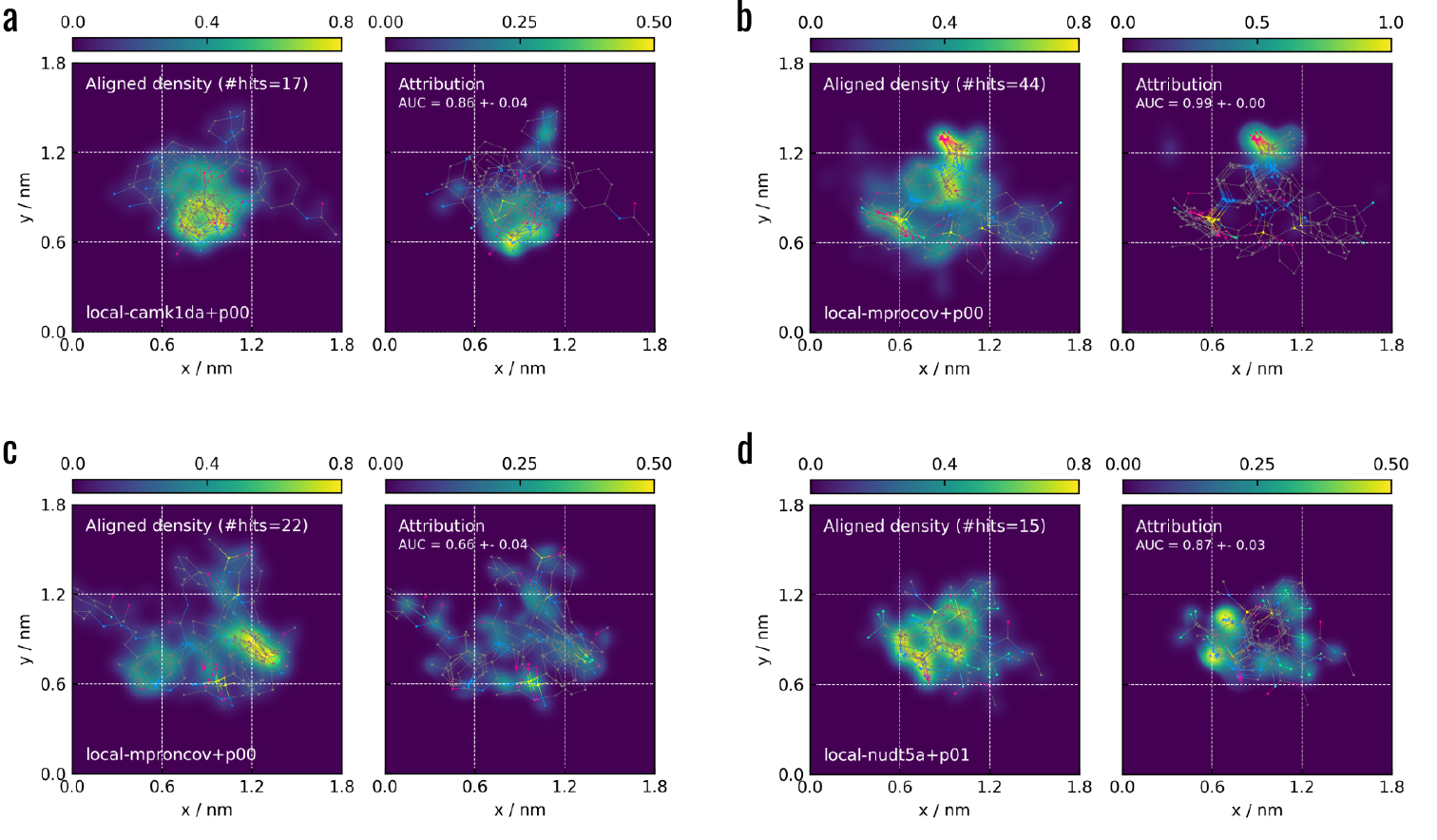}
\caption{Superimposed ligand densities and machine-learned attribution fields derived from a subset of the XChem data. The systems shown are (a) Calcium-/calmodulin-dependent kinase 1Da, (b) SARS-CoV-2 main protease (covalent hits), (c) SARS-CoV-2 main protease (non-covalent hits), (d) ADP-sugar pyrophosphatase. In each panel, the superimposed density is shown on the left, the mapped attribution field on the right. }
\label{fig:xchem}
\end{figure*}

\begin{figure*}[t]
\centering
\includegraphics[width=0.7\linewidth]{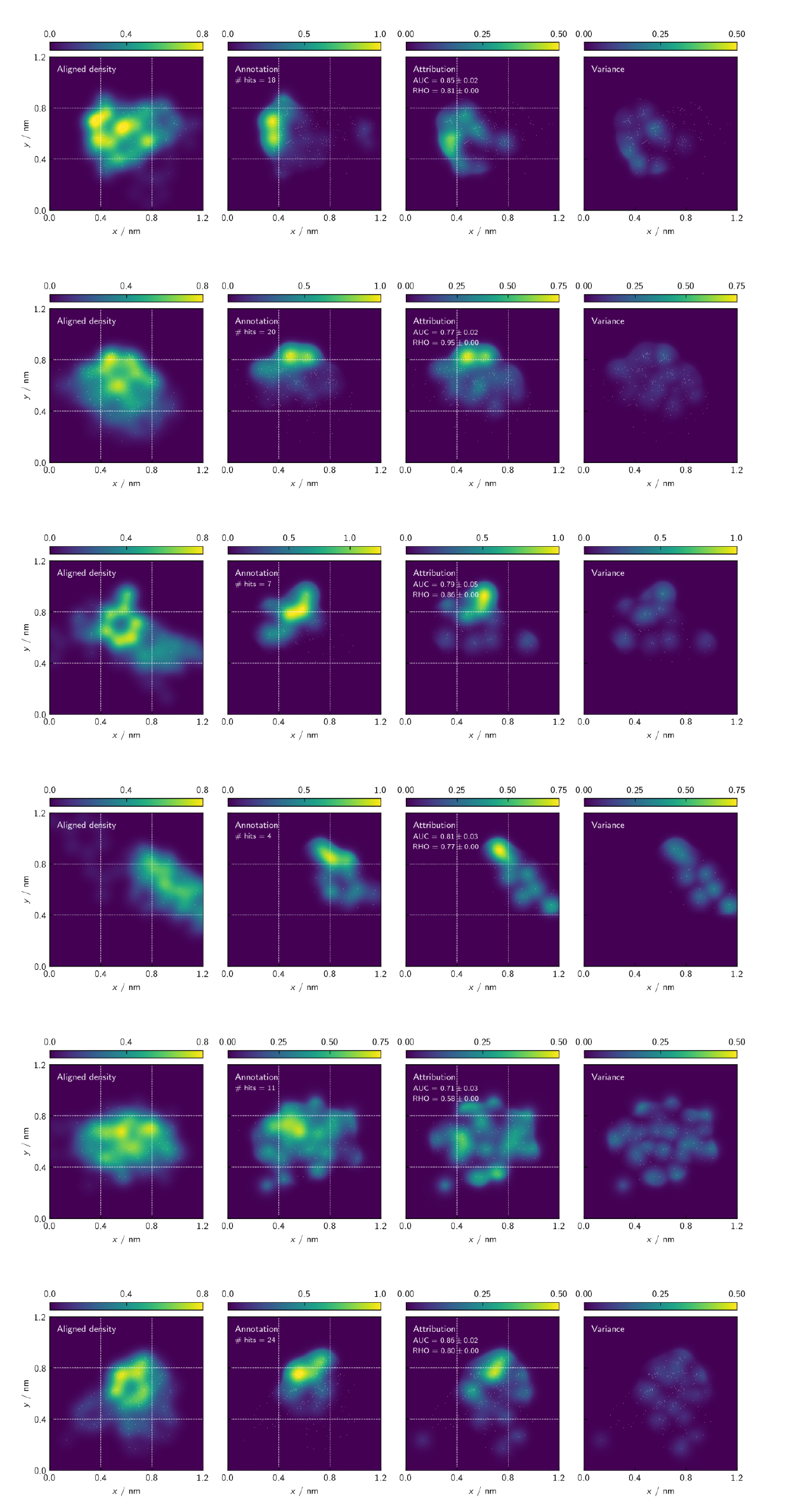}
\caption{ (1/3) Comparison between manual and machine-learned annotations of superimposed fragment hits. Each row consists of: left, the superimposed ligand density; centre-left, the manually assigned pharmacophore weights mapped onto this density; centre-right, the attribution field derived from the filtered weights; right, the standard deviation of the predicted attribution field.}
\label{fig:sijuxta_0}
\end{figure*}

\begin{figure*}[t]
\centering
\includegraphics[width=0.7\linewidth]{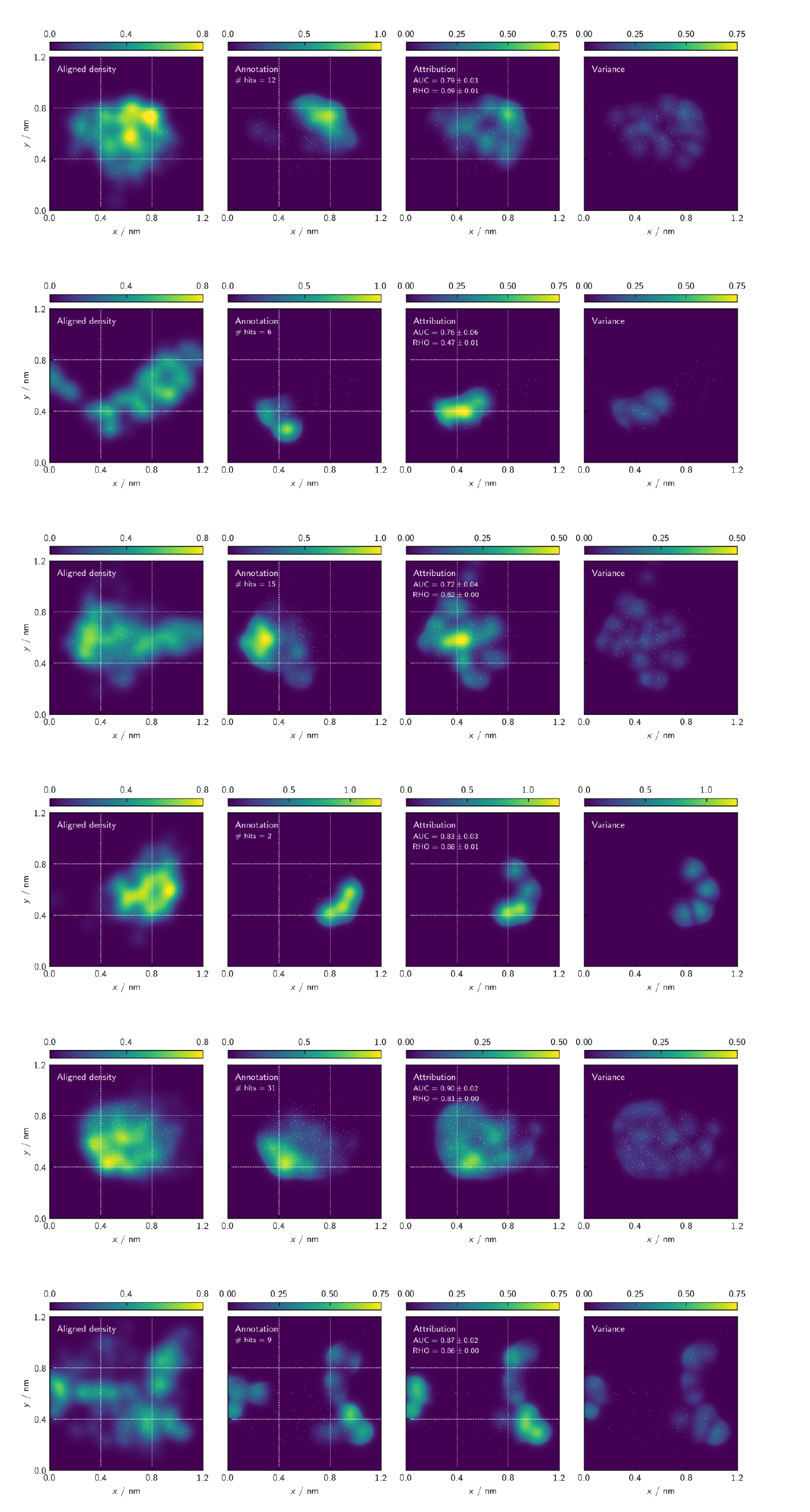}
\caption{ (continued, 2/3) Comparison between manual and machine-learned annotations of superimposed fragment hits. Each row consists of: left, the superimposed ligand density; centre-left, the manually assigned pharmacophore weights mapped onto this density; centre-right, the attribution field derived from the filtered weights; right, the standard deviation of the predicted attribution field.}
\label{fig:sijuxta_1}
\end{figure*}

\begin{figure*}[t]
\centering
\includegraphics[width=0.7\linewidth]{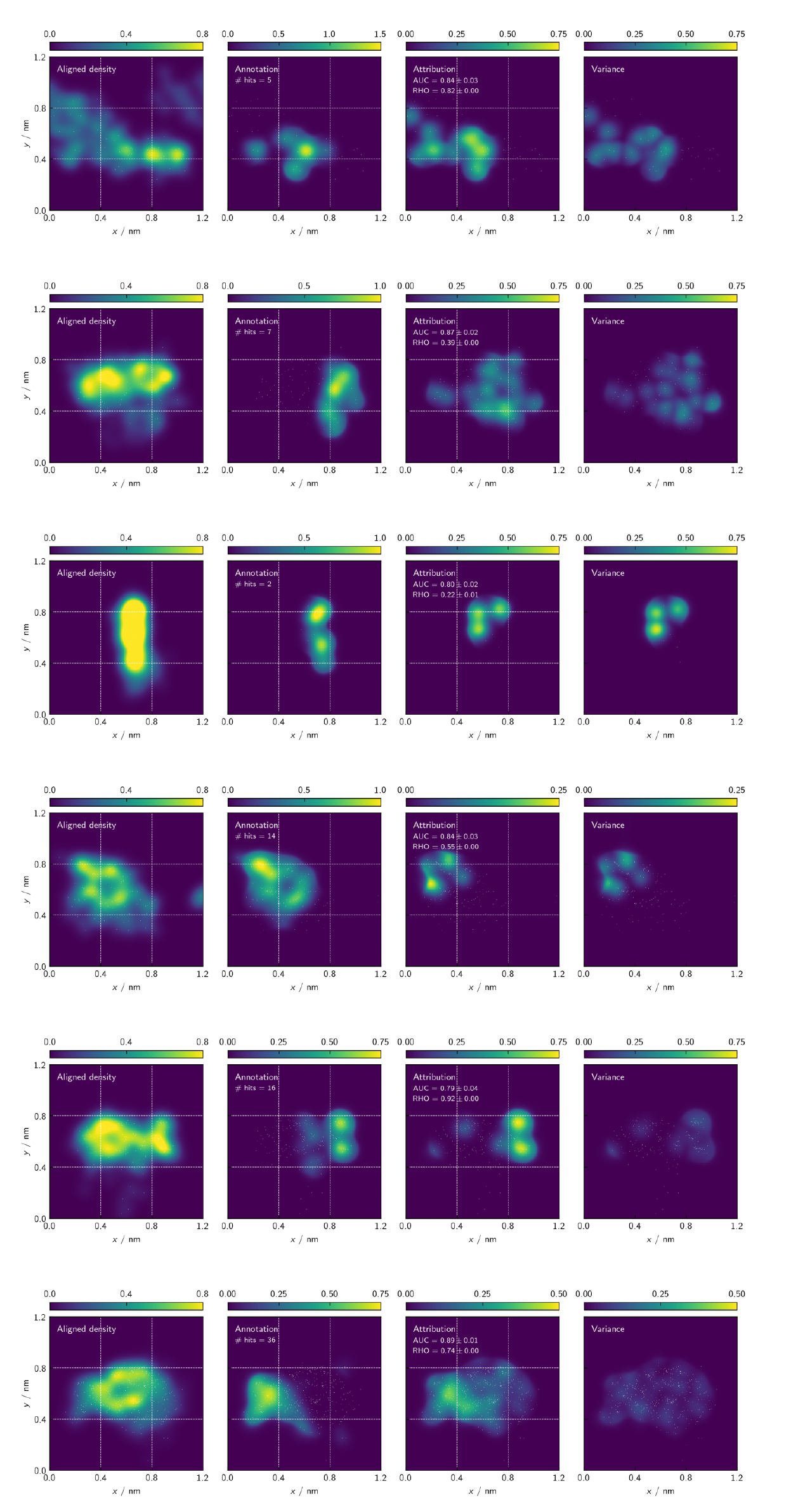}
\caption{ (continued, 3/3) Comparison between manual and machine-learned annotations of superimposed fragment hits. Each row consists of: left, the superimposed ligand density; centre-left, the manually assigned pharmacophore weights mapped onto this density; centre-right, the attribution field derived from the filtered weights; right, the standard deviation of the predicted attribution field.}
\label{fig:sijuxta_2}
\end{figure*}


\end{document}